\documentclass[prb,aps,twocolumn,showpacs]{revtex4}

\usepackage{graphicx}
\usepackage[usenames]{color}
\definecolor{orange}{rgb}{0.7,.35,0}

\def\ot{\frac{1}{3}}
\def\tf{\frac{2}{5}}
\def\tt{\frac{2}{3}}

\def\up{\uparrow}
\def\dn{\downarrow}
\def\ve{\varepsilon}

\def\cunit{{e^2}/{4\pi\ve \ell_0}}

\def\vek#1{\vec{#1}}
\def\ket#1{|#1\rangle}

\def\krv{\vek{k}_r}
\def\etal{{\em et al.}}

\begin{document}

\title{Integral and fractional Quantum Hall Ising ferromagnets}

\author{
   Karel V\'yborn\'y,$^{1,2}$
   Ond\v rej \v Cert\'\i k,$^1$
   Daniela Pfannkuche,$^2$
   Daniel Wodzi\'nski,$^{3,4}$
   Arkadiusz W\'ojs,$^{3,4}$ and
   John J. Quinn$^4$}

\affiliation{
   \mbox{
   $^1$ Fyzik\'aln\'\i{} \'ustav Akademie v\v ed \v CR
       Cukrovarnick\'a 10, 162 53 Praha 6, Czech Republic}\\
   $^2$I. Institut f\"ur Theoretische Physik, Universit\"at Hamburg,
       Jungiusstrasse 9, 20355 Hamburg, Germany\\
   \mbox{
   $^3$Institute of Physics, Wroc{\l}aw University of Technology, 
       Wybrze\.ze Wyspia\'nskiego 27, 50-370 Wroclaw, Poland}\\
   $^4$Department of Physics, University of Tennessee, 
       Knoxville, Tennessee 37996, USA}

\begin{abstract}
\end{abstract}
\pacs{73.43.-f,71.10.Pm,75.10.Jm}
\date{September 29th, 2006}

\begin{abstract}
We compare quantum Hall systems at filling factors $\nu=2$ to
$\nu={2\over3}$ and ${2\over5}$, corresponding to the exact filling of
two lowest electron or composite fermion (CF) Landau levels.  The two
fractional states are examples of CF liquids with spin dynamics.
There is a close analogy between the ferromagnetic (spin polarization
$P=1$) and paramagnetic ($P=0$) incompressible ground states that
occur in all three systems in the limits of large and small Zeeman
spin splitting.  However, the excitation spectra are different.  At
$\nu=2$, we find spin domains at half-polarization ($P={1\over2}$),
while antiferromagnetic order seems most favorable in the CF systems.
The transition between $P=0$ and 1, as seen when e.g. the magnetic
field is tilted, is also studied by exact diagonalization in toroidal
and spherical geometries.  The essential role of an effective CF--CF
interaction is discussed, and the experimentally observed
incompresible half-polarized state is found in some models.

\end{abstract}

\maketitle

\section{Introduction}

The long range spin order in quantum Hall systems at integer filling
factor can easily be explained in terms of single electron Landau
levels and it became common to call these systems quantum Hall
ferromagnets\cite{jungwirth:12:2000} (QHF). Electron-electron
repulsive interaction, in particular exchange, is known to further
stabilize the ferromagnetism. Given the composite fermion (CF)
mapping\cite{Jain89}, it was not surprising that ground states (GS)
with spin order were found also at fractional filling factors which
correspond to integer filling factor of CFs. However, since the
interactions between CFs are different to the previous case, it was
not clear whether they promote or suppress the
ferromagnetism. Moreover, the interaction can be more important in the
fractional regime, and it might even destroy the ferromagnetism,
especially near a transition between two GS of different order,
because it cannot be downscaled as in the integer regime for
$B\to\infty$. The field of fractional QHF became particularly
interesting when signatures of states with intermediate polarization
were experimentally discovered near the transition.

The mentioned GS transition in systems at filling factor
$\nu=n/(eB/h)=2$ occurs when the $0$$\up$ and $1$$\dn$ Landau levels (LL)
cross \cite{Giuliani84,wojs:04:2002}. This happens when we vary
the ratio of Zeeman and cyclotron energies ($\theta$) as it is the
case with tilting the magnetic field $B$, changing the $g$--factor or
pumping the nuclear spins of the host lattice.  The $0$$\dn$ LL is
always full at filling factor two, while $1$$\dn$ and $0$$\up$ is full 
and empty for large $\theta$ and vice versa for small $\theta$. The GS is
thus fully spin polarized in the first case and it is a spin singlet
in the second case. The stabilizing effect of interactions implies
that the transition between these two spin--ordered states is abrupt
without any intermediate state when $\theta$ is varied.

The situation is different at filling factors $\nu=\tf$ and $\tt$
which both correspond to filling factor $\nu^*=2$ of composite
fermions (in the latter case, the effective magnetic field acting on
CFs points in opposite direction to $B$). The crossing of $0$$\up$ and
$1$$\dn$ CF LLs is now induced by varying the ratio of Zeeman and
Coulomb energy $\eta\propto \sqrt{B}$, since the CF cyclotron energy
($\hbar\omega_c^*$) is determined fully by the electron--electron
interaction if LL mixing is neglected. Optical experiments by
Kukushkin \etal\cite{kukushkin:05:1999} confirmed the transition from
$P=0$ to $P=1$ when $\eta$ was increased, but they also revealed a
stable intermediate state at $P=0.5$. Experiments by Freytag
\etal\cite{freytag:09:2001} suggested another intermediate state with
$P\approx 0.8$. On the other hand, transport
measurements\cite{kraus:12:2002,kronmuller:09:1998,%
hashimoto:04:2002,smet:01:2002,smet:03:2001} showed huge longitudinal
magnetoresistance at the transition, which was attributed to domain
formation as an opposite to a homogeneous incompressible quantum Hall
state.

Possible stable half--polarized states in the context of $\nu=\tf$ and
$\tt$ were then discussed by Apal'kov \etal\cite{apalkov:02:2001}
(condensate of $L=1$ excitons), Murthy\cite{murthy:01:2000} using the
Hamiltonian theory\cite{murthy:10:2003} of CF (quantum Hall crystals),
Mariani \etal\cite{mariani:12:2002} and Merlo
\etal\cite{merlo:04:2005} (pairing of CFs similar to
superconductivity) and more recently also by Yang
\etal\cite{yang:08:2005} (unidirectional CDW of CF). Spin transitions
and instabilities were also studied in other QH systems including
$\nu=2$ (Giuliani \etal\cite{Giuliani84}), $\nu=4/3$ (one of the
authors\cite{wojs:04:2002}) or higher integer fillings (Rezayi
\etal\cite{rezayi:05:2003}).

The purpose of this article is primarily to compare systems of
electrons and composite fermions at the same filling factor $\nu$ or
$\nu^*$ equal to two. We numerically investigate the {\em
electrons} at filling $\nu=2$ versus $\nu=\tf$ and $\tt$. We show that
even though the ground states are analogous, the excitations are quite
different. As a consequence, physics of the paramagnet-ferromagnet
transition is distinct in the two systems. A low-energy half-polarized state
with antiferromagnetic spin order is found at $\nu=\tt$, while domains
of $P=0$ and $P=1$ are found at $\nu=2$.  Next we turn
to the concept of composite fermions (CF). It is demonstrated that the
effective interaction between the CFs calculated near $\nu=\ot$ (or
$\nu^*=1$) leads to very questionable results when applied to
$\nu=\tf$, $\tt$ and we discuss alternative approaches.

\subsection{Exact diagonalization}

Each Landau level (LL) is highly degenerate, in a given area it can accommodate
(maximum of) $N_m$ electrons of exactly the same energy. When studying
the quantum Hall ferromagnets (QHF), it happens frequently that $N$
particles, electrons or composite fermions, are to occupy these states.
In absence of interaction this yields a vast number of degenerate
$N$--particle states: ${N_m\choose N}$. With particle--particle
interaction switched on, no perturbation theory is tractable, as it
requires a unique ground state to start with. The standard way to
handle this problem is to diagonalize the full Hamiltonian with
respect to the full basis of dimension $({N_m\atop N})$. The
Hamiltonian comprises primarily of the particle--particle interaction
and the fundamental approximation made is that we consider a finite
system of $N$ particles rather than an infinite one.

Within Haldane's model\cite{Haldane83},
$N$ electrons are confined to a spherical surface of radius $R$.
Dirac monopole of strength $2Q$ (in the units of elementary flux 
quantum, $hc/e$) in the center acts as a source of radial (i.e., 
normal to the surface) magnetic field $B$. 
Magnetic length scale $\ell_0=\sqrt{hc/eB}$ is simply related 
to $2Q$ and $R$ by $R^2=Q\ell_0^2$.

The lowest (0th) Landau level (LL) is a shell of angular momentum 
$l_0=Q$ and finite degeneracy $g_0=2l_0+1$ (with different orbitals
distinguished by angular momentum projection $m$).
Higher LL's, labeled by $n>0$, have $l_n=Q+n$ and $g_n=2l_n+1$.
Including spin, the single-particle states on Haldane sphere 
(called ``monopole harmonics'') \cite{Wu76} are uniquely denoted 
by $i=[n,m,\sigma]$.
The dependence of LL degeneracy $g_n$ on the LL index $n$ is a 
finite-size artefact of spherical geometry, known to cause some 
inconvenience in calculations involving different LL's. 

The integration of two-body interaction (Coulomb) matrix elements
$\left<i,j|V|k,l\right>$ can be done analytically for an ideally 
2D system \cite{Fano86}.
For finite width $w$ of the electron layer, a fixed density profile
$\varrho_w(z)$ can be used to model the lowest subband in the normal 
direction, and the calculation of $\left<i,j|V|k,l\right>$ involves 
one-dimensional numerical integration. 
These two-body matrix elements are related with Haldane interaction 
pseudopotential (pair interaction energy as a function of relative
angular momentum $\mathcal{R}$) \cite{Haldane87} through the 
Clebsch-Gordan coefficients.

A complication with the spherical geometry is the definition of the
filling factor: $\nu=N/(2l_0+\gamma)$. The ``shift'' $\gamma$ is a
topological quantum number\cite{nayak:07:1995} which is of the order
of one, independent on $N$ but it need not be the same for different
states at the same value of $\nu$. Thus, looking for one particular
state at a given filling factor, we must also know its $\gamma$ and
adjust the value of $l_0=Q$ properly (simple $2l_0=N/\nu$ may not
work). Further implications of this fact are discussed below
(Sec.~\ref{pos-01}).

The torus geometry\cite{yoshioka:04:1983,yoshioka:06:1984}, or
rectangle ($a$ by $b$) with periodic boundary
conditions\cite{haldane:02:1985} (PBC), is characterized by the number of
single particle states $N_m$ and aspect ratio $\alpha=a:b$. The area
of the rectangle is fixed by $ab=2\pi\ell_0^2 N_m$
\cite{chakraborty:1995}. The filling factor is $\nu=N/N_m$ when
$N$ electrons are put into the rectangle. The Haldane
pseudopotentials can be defined in this geometry, too, albeit they no
longer correspond to eigenstates of angular
momentum\cite{vyborny:2005}.

The rotational symmetry of a sphere, implying the angular momentum
$|\vek{L}|$ and its $z$--component $L_z$ to be good quantum numbers,
is replaced by the invariance to magnetic translations in the
rectangle with PBC (described in detail by
Haldane\cite{haldane:11:1985}). The corresponding good quantum numbers
are linear momentum along the sides of the rectangle, $k_x$ and
$k_y$. These can take on discrete values\cite{comm:02} $k_x=i k_u$ and $k_y=j
k_u\alpha$ with $i,j=0,\pm 1,\ldots,\pm [N/2]$ and
$k_u=\sqrt{2\pi/(N_m\alpha)}/\ell_0$ (note that this depends on the
filling factor\cite{haldane:11:1985}). The magnetic Brillouin zone is
therefore rectangular and its size is grows with system size ($\propto
\sqrt{N_m}$; Fig.~6 in Ref. \cite{vyborny:2005}).

Conceptually, $|\vek{L}|$ and $L_z$ of the sphere correspond to
$|\vek{k}|$ and $k_y$ on the torus. Indeed, the ED spectra from both
geometries mapped using $|\vek{L}|/\hbar=|\vek{k}| R$ are in a good
quantitative agreement. However, the representations of these
symmetries do differ in {\em finite} systems. The relationship between
the orbital degeneracy of a given level and its $|\vek{k}|/k_u$ is
non-trivial (nonmonotonous and $N_m$--dependent), while there are
always $2L+1$ degenerate states for a level with total angular
momentum $L$. Moreover, the orbital degeneracy on the torus,
corresponding to rotational symmetry in an infinite system, can easily
be lifted by displacing the aspect ratio $\alpha$ slightly from one.
We may expect that isotropic states (e.g. a single quasiparticle on the
background of an isotropic ground state) will suffer less from the
finite size when studied on a sphere. Translationally invariant but
anisotropic states (such as a plane wave) will be better served on a torus.

\subsection{Quantum Hall ferromagnets}

The basic fact about quantum Hall systems is that in a situation where
$N$ electrons have the freedom to occupy $N$ places
(single--electron orbitals) out of $2N$, the ground state will
be unique and it will possess long--range spin order. 

Such a situation typically occurs when two Landau levels are
degenerate. The best known example is $\nu=1$ at zero Zeeman
energy. Here, $N=eB/h$ electrons (per unit area) can choose any of
$2eB/h$ single-electron states available in the degenerate $0$$\up$
and $0$$\dn$ Landau levels. The electron-electron interaction
implies\cite{girvin:07:1999} a unique ground state $\ket{\Psi}$ which
is the completely filled $0$$\up$ LL or any state $R\ket{\Psi}$ where
$R$ is an arbitrary rotation of the total spin [leading to an $SU(2)$
symmetry of the GS]. The popular explanation of this effect is the
tendency to maximize the gain in Coulomb exchange energy.  All spins
in the ground state must be parallel to each other but the direction
can be arbitrary.  This renders the $\nu=1$ system to be called a
Heisenberg ferromagnet.

Here we investigate another system. A different QHF occurs at $\nu=2$ when
$0$$\dn$ and $1$$\up$ LL are degenerate as it is the case when the
cyclotron energy is equal to the Zeeman splitting. The low lying
$0$$\up$ LL is completely filled and it can be considered inert. The two
crossing levels then again dispose of $2eB/h$ single-electron states
to be occupied by $eB/h$ electrons. This time, the ground state is
twofold degenerate and it consists either of the completely occupied
$0$$\dn$ LL or the completely occupied $1$$\up$ LL (disregarding the
occupied $0$$\up$ LL). The $Z_2$ symmetry of the GS, regarding the inversion of
all spins in the active LLs, earned this system the name Ising QHF.
 
To make the $0$$\dn$ and $1$$\up$ LL degenerate, the Zeeman energy has
to be adjusted properly to compensate the difference in their
Hartree--Fock selfenergies $\Sigma_{n\sigma}$ (unlike
$\Sigma_{0\downarrow}$, the selfenergy $\Sigma_{1\uparrow}$ includes
exchange with the completely filled $0\uparrow$ LL in addition to
the cyclotron energy). Throughout this article we will almost always
work at the degeneracy, hence Zeeman energy will be included making energies
of the two $Z_2$ symmetric ground states ($S=0$ and $S=N/2$,
when the occupied $0$$\up$ LL is included) equal. Experimental
techniques to achieve this situation (tilted magnetic field,
$g$--factor reduced by hydrostatic pressure etc.) are summarized
elsewhere\cite{chakraborty:07:2000,vyborny:2005}.

\section{Polarizations from zero to one}

\subsection{Calculation in terms of electrons}
\label{pos-01}

To study the transition between polarized and unpolarized $\nu^*=2$ 
($\nu=2/3$ and $2/5$) Jain states on a sphere, we have used the series 
of finite systems $(N,2Q)$ in which these ground states occur and 
compared their energy spectra for arbitrary spin configurations.
The relation between $N$ and $2Q$ for these Jain states is obtained 
from the condition of complete filling of two lowest composite fermion 
(CF) shells at the effective magnetic monopole strength $2Q^*=2Q-2(N-1)$.
The $\nu=2/5$ and $2/3$ states occur for $2Q^*$ having the same or 
opposite sign to $2Q$, respectively \cite{Heinonen98,Yoshioka02}.

Because $g_0^*=2Q^*+1$ is different from $g_1^*=2Q^*+3$, the polarized
and unpolarized $\nu^*=2$ states occur in different systems $(N,2Q)$,
corresponding to $N=g_0^*+g_1^*$ or $2g_0^*$, respectively.  As an
unfortunate artefact of the spherical geometry, the ``shift'' $\gamma$
is thus different for the polarized and unpolarized states of
electrons, namely $\gamma=4$ and 3 (at $\nu=2/5$) and $\gamma=0$ and 1
(at $\nu=2/3$).

The fact that the pair of polarized and unpolarized $N$-electron 
$\nu^*=2$ (either $\nu=2/5$ or $2/3$) states do not occur at the 
same value of $2Q$ prevents transition (on a sphere) from one to 
the other through a sequence of spin-flips.
Beginning from a polarized state and flipping consecutive $K=N/2-1$ 
spins leads to the system containing two CF's in the $n^*=1$ LL 
(so-called quasielectrons, QE's) in addition to the unpolarized 
state of $N-2$ electrons.
On the other hand, beginning from an unpolarized state and flipping 
consecutive $K=N/2$ spins leads to the system containing two CF 
vacances (called quasiholes, QH's) in the polarized state of $N+2$ 
electrons.

This discrepancy complicates calculation of the ground state energy 
$E$ at a fixed filling factor ($\nu=2/5$ or $2/3$) as a function of 
spin polarization $P=(N_\uparrow-N_\downarrow)/(N_\uparrow+N_\downarrow)
=2S_z/N$.
In contrast to torus geometry, it cannot be simply calculated as 
the ground state energy for fixed $(N,2Q)$ as a function of $S_z$.
This forces one into comparison of energies obtained for different 
$N$ or $2Q$ (problematic in small systems because the energies of 
the QE's and QH's and of the underlying incompressible Jain state 
scale differently with $N$, and because the surface curvature 
$R^{-1}$ affecting all interaction energies depends on $2Q$).

Therefore, we have calculated separately the energy as a function 
of $S_z$ for only up to a few spin flips away from the polarized 
and from the unpolarized $\nu^*=2$ state (this corresponds to 
studying the behavior of $E(P)$ separately at $P<1/2$ and $P>1/2$, 
and leaving the $P\sim1/2$ regime unknown).
Recall, however, that the goal is to find the $E(P)$ curve at the 
Zeeman energy $E_Z=E_Z'$ for which the polarized and unpolarized 
Jain states are degenerate, $E(0)=E(1)$.
Hence, in order to estimate $E_Z'$, one has to know the energies 
of both polarized and unpolarized Jain states corresponding to the 
same $(N,2Q)$.
In other words, one needs the estimate of $E/N$ (energy per electron)
in both polarized and unpolarized Jain state in small systems.

\begin{figure}
\resizebox{3.4in}{1.75in}{\includegraphics{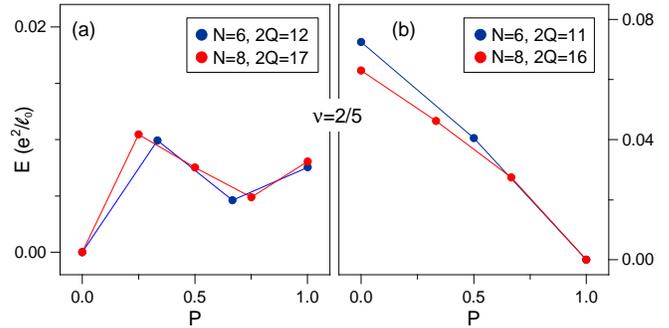}}
\caption{Excitation energy $E$ as a function of electron spin polarization 
   $P$, calculated on a sphere for $N=6$ and 8 electrons, at the 
   values of $2Q$ corresponding to incompressible $\nu=2/5$ Jain 
   states at $P=0$ (a) or $P=1$ (b).}
\label{fig-31}
\end{figure}

\begin{figure}
\resizebox{3.4in}{1.75in}{\includegraphics{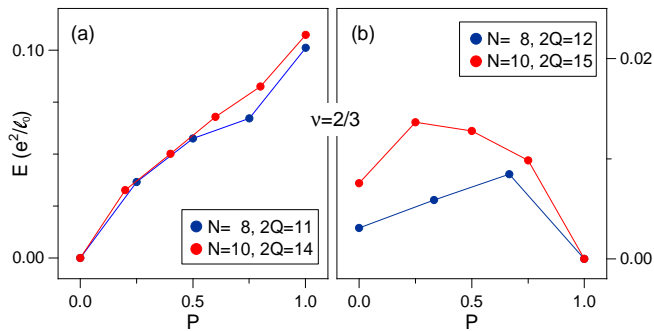}}
\caption{Same as Fig.~\ref{fig-31}, but $\nu=2/3$ and $N=8, 10$.} 
\label{fig-32}
\end{figure}

\begin{figure}
\begin{tabular}{cc}
\includegraphics[scale=0.5]{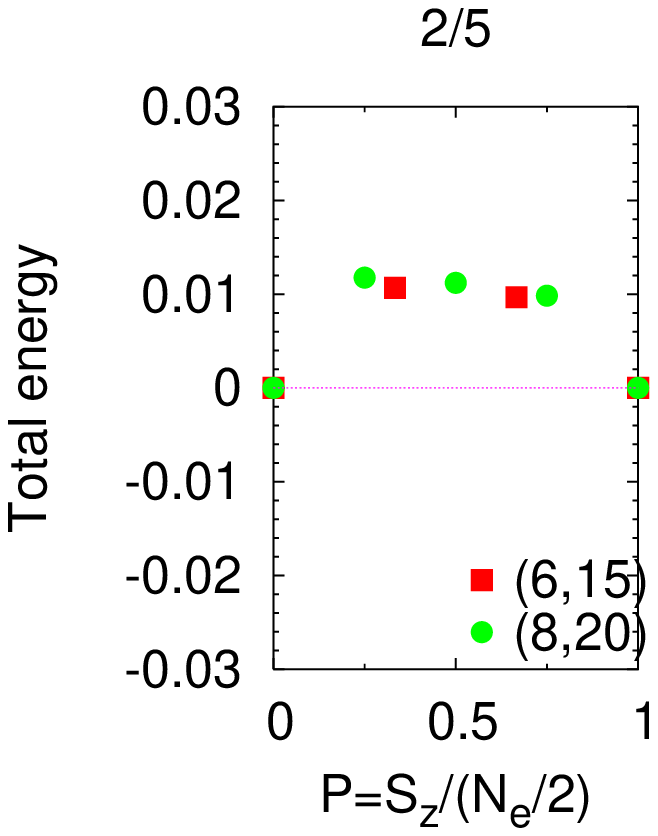} &
\includegraphics[scale=0.5]{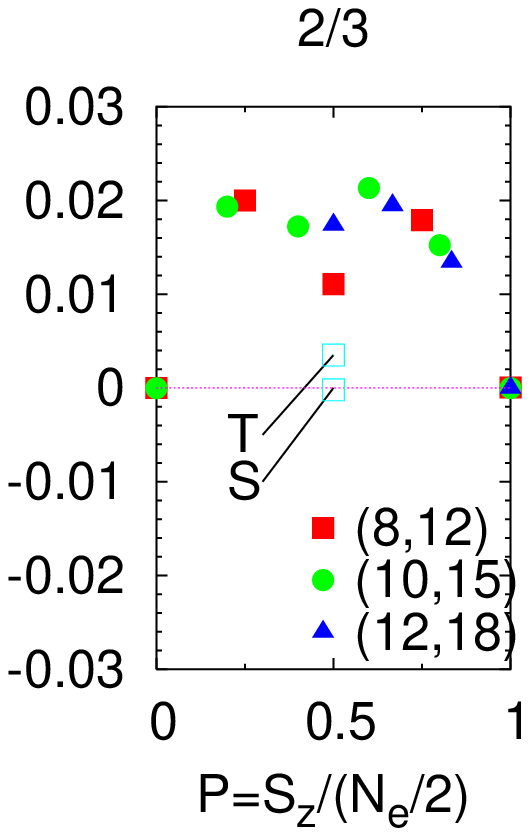} \\
(a) & (b)\\
\end{tabular}
\caption{$E(P)$ of the $\tf$ and $\tt$ systems on a torus. Here, the
  singlet and polarized states occur in systems with the same
  $(N,N_m)$, so that the $E(P)$ dependence shown here is relevant in
  the whole range of $P$. The $1/N\to 0$ extrapolated values are
  displayed by 'S' (sphere) and 'T' (torus).
}
\label{fig-01}
\end{figure}

\begin{figure}
\begin{center}
\includegraphics[scale=0.6]{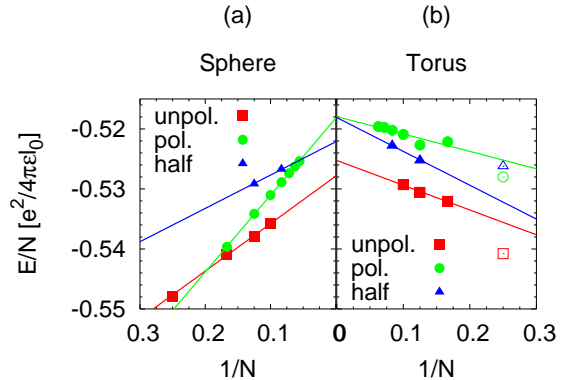}
\end{center}
\caption{Energy per particle of the singlet state, the HPS and the
  fully polarized state at $\nu=\tt$. Different system sizes are shown
  ($N=4$ to 18 electrons), extrapolation to infinite systems,
  $1/N\to 0$ (only solid points were used).}
\label{fig-03}
\end{figure}

Let us explain how it is done on the example of $\nu=2/5$ and $N=8$.
To get $E(P)$ at small $P$ we use finite-size calculation of $E(S_z)$ 
starting with the unpolarized Jain state at $2Q=17$.
In the CF picture of this state, $2Q^*=3$, $g_0^*=4$, and the $N=8$ 
CF's fill completely the $0$$\uparrow$ and $0$$\downarrow$ LL's.
We calculate $E(S_z)$ in the whole range of $S_z$.
At $S_z=N/2$, the ground state contains two QH's in the polarized 
Jain state, whose energy must be subtracted to find $E_Z'$ correctly.
This is done easily by replacing the calculated $E(S_z=N/2)$ by the 
energy obtained for the polarized Jain state at $2Q=16$, rescaled
appropriately by $\sqrt{16/17}$ to account for a different $\ell_0$.
With such estimate of $E_Z'$ we ignore all but the few values of
$E(S_z)$ corresponding to the smallest $S_z$, and recalculate them
into $E(P)$ at small $P$.
The result is plotted in Fig.~\ref{fig-31}a, also showing weak 
finite-size effects (due to the $g_0^*\ne g_1^*$ discrepancy) 
at small $P$, confirmed by comparison with the $N=6$ calculation.
The calculation of $E(P)$ at large $P$ goes analogously, starting
with the polarized Jain state at $2Q=16$, and with the energy of the 
unpolarized Jain state needed for $E_Z'$ obtained by rescaling the 
value at $2Q=17$ and $S_z=0$.
Again, as shown in Fig.~\ref{fig-31}b, comparison of data for $N=6$ 
and 8 confirms the size-convergence.
The same procedure has been carried out for $\nu=2/3$, with the
results plotted in Fig.~\ref{fig-32}a,b.

While the calculation on a sphere allowed us to find $E(P)$ only 
at small or large $P$ (and with some uncertainty in the estimate 
of $E_Z'$), one conclusion seems established despite finite-size 
problems: $E(P)$ at $E_Z'$ increases when $P$ is either increased
from 0 or decreased from 1. 
In other words, $E(P)$ at small either $P$ or $1-P$ is larger 
than $E(0)=E(1)$, which would imply abrupt transition between 
the polarized and unpolarized $n^*=2$ Jain states as a function 
of $E_Z$.

Contrary to the sphere, it is possible to scan the whole range of $P$
using calculations on a torus (square with periodic boundary
conditions). Choosing $(N,N_m)$, the number of electrons and
magnetic flux quanta, the sector of spin $S_z$ corresponds to
filling factor $\nu=N/N_m$ and polarization $P=S_z/(N/2)$. 

The principal conclusion is to confirm the observations made for a
sphere: all states with intermediate polarization $0<P<1$ are higher
in energy than the polarized ($P=1$) and the singlet state
($P=0$). Systems of different sizes exhibit qualitatively the same
behaviour both at filling factor $\tf$ and $\tt$,
Fig.~\ref{fig-01}. On a quantitative level and within the small
systems accessible to exact diagonalization, it seems unlikely that
this situation changes if we continue to larger systems. Nevertheless,
the ground state energies at $P=0,\frac{1}{2}$ and $1$ extrapolated to
$1/N\to 0$, Fig.~\ref{fig-03}, give opposite predictions in this
aspect: $E(\frac{1}{2})>E(0)=E(1)$ on a torus while the oposite was
found on the sphere\cite{niemela:xx:2000}. It is, however, important
to note that the extrapolation for $E(\frac{1}{2})$ is based only on
two points. 
Moreover, it cannot be established reliably whether all
these points correspond to a realisation of the same state in an
infinite system\cite{vyborny:2005}. On the other hand, it is possible
to speculate that the $P=\frac{1}{2}$ state of $\nu=\tt$ is more
stable than other states with $|P-\frac{1}{2}|<\frac{1}{2}$, i.e. that
there may be a downward cusp at $P=\frac{1}{2}$, Fig.~\ref{fig-01}b.

\subsection{Composite fermion calculation}

In the composite fermion (CF) picture \cite{Jain89}, the $\nu=2/3$ 
and $2/5$ fillings correspond to the same effective CF filling factor 
$\nu^*=2$.
The completely polarized ($S=N/2$ and $P=1$) and unpolarized ($S=P=0$) 
states are represented by simple single-particle configurations, 
with a pair of completely filled CF LL's: $0$$\uparrow$ and either 
$1$$\uparrow$ or $0$$\downarrow$.
Stability of the corresponding four Jain states ($\nu=2/3$ or $2/5$; 
$P=0$ or 1) requires a splitting $\Delta$ of the $1$$\uparrow$ and 
$0$$\downarrow$ LL's, and their incompressibility is attributed to 
a single-particle (cyclotron) CF gap $\hbar\omega_c^*$.

The intermediate polarizations ($0<P<1$) are possible for nearly
degenerate CF self-energies $\Sigma_{n\sigma}$ of the $1$$\uparrow$
and $0$$\downarrow$ LL's (Sec. I.B).  Whether such partially polarized
states indeed occur for some appropriate $\Delta$ depends on the
CF--CF interactions within and between the two partially filled CF
LL's.  If they do, their many-CF wavefunctions, the CF--CF
correlations, and possible incompressibility (at least at some of the
intermediate values of $P$) also depends completely on the CF--CF
interaction.  Remarkably, the $P=1$ to 0 (paramagnet-ferromagnet)
transition occurs directly for the electrons filling two LL's
($\nu=2$) \cite{Giuliani84}, but partially polarized states were
suggested in a mixed electron--CF system at $\nu=4/3$ \cite{wojs:04:2002}.

Interaction between two particles (e.g., electrons or CF's) in a pair
of LL's $(n_1,\sigma_1)$ and $(n_2,\sigma_2)$ is determined by Haldane 
pseudopotential $V(\mathcal{R})$, defined as pair interaction energy 
as a function of relative angular momentum \cite{Haldane87}.
For a pair of identical CF's in the same LL (here, $0$$\downarrow$ 
or $1$$\uparrow$), the allowed $\mathcal{R}$'s are odd integers.
For two CF's distinguished by spin and/or LL index, $\mathcal{R}$ 
can be odd or even.
Assigning pseudospins $\uparrow$ and $\downarrow$ to the CF LL's 
$1$$\uparrow$ and $0$$\downarrow$, the CF dynamics within these 
two levels is determined by a set of three pseudopotentials:
$V_{\uparrow\uparrow}(\mathcal{R})$ and $V_{\downarrow\downarrow}
(\mathcal{R})$ for $\mathcal{R}=1$, 3, 5, \dots, and $V_{\uparrow
\downarrow}(\mathcal{R})$ for $\mathcal{R}=0$, 1, 2, \dots.
Because $V_{\uparrow\uparrow}$ and $V_{\downarrow\downarrow}$ 
describe interaction in different LL's, they are not equal.
Hence, the CF--CF interaction within these two LL's is 
pseudospin-asymmetric.

Knowing $V$ is the key to understanding the CF dynamics in partially
filled shells.
The correlations have particularly simple form when $V(\mathcal{R})$ 
is dominated by one coefficient.
In this case, the particles interacting through $V(\mathcal{R})$ tend 
to avoid the corresponding high-energy pair state \cite{Haldane87}.
More generally, the correlations depend only on the anharmonic
contribution to $V(\mathcal{R})$, where the harmonic dependence means 
$V$ linear in average squared distance $\left<r^2\right>$, which 
corresponds to a roughly linear $V(\mathcal{R})$ \cite{Wojs00a}.

\begin{figure}
\resizebox{3.4in}{1.75in}{\includegraphics{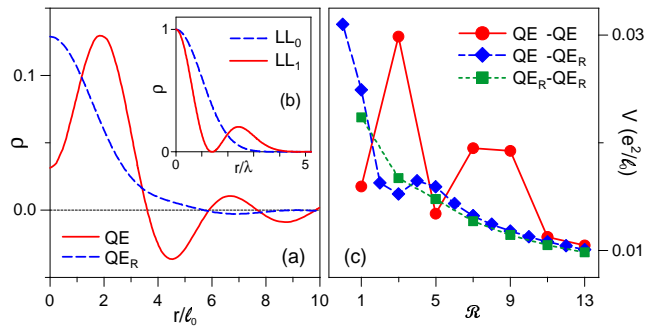}}
\caption{
   (a) Radial charge distributions of QE$_{\rm R}$ and QE 
   at $\nu=1/3$, i.e., of the CF's in the lowest and first 
   excited LL.
   (b) Same for the electrons.
   (c) CF--CF interaction pseudopotentials at $\nu=1/3$.}
\label{fig-11}
\end{figure}

The three CF--CF pseudopotentials, $V_{\uparrow\uparrow}$,
$V_{\downarrow\downarrow}$, and $V_{\uparrow\downarrow}$, are known 
quite well for electron systems near $\nu=1/3$, where they describe 
interactions among Laughlin QE's \cite{Laughlin83} or reversed-spin 
QE$_{\rm R}$'s \cite{Chakraborty86,Rezayi87}.
At long range (large $\mathcal{R}$) these pseudopotentials must be
consistent with Coulomb repulsion of two fractional charges, $-e/3$.
At short range they can be obtained from finite-size calculations
\cite{Sitko96,Wojs00b,Lee01,Szlufarska01} and show features revealing 
the CF internal structure.
Radial charge distributions of QE and QE$_{\rm R}$'s are presented 
in Fig.~\ref{fig-11}a.
They were calculated numerically from the exact eigenstates of 10 
electrons, and they are normalized to $\int\varrho(r)rdr=1/3$ (in 
the length units of $\ell_0$).
Comparison with the electron charge profiles plotted in the inset 
shows that, except for the reduced QE/QE$_{\rm R}$ charge, the CF's 
and electrons in their lowest LL's are very similar 
(QE$_{\rm R}$ and LL$_0$ in Fig.~\ref{fig-11}a,b), while the CF's
and electrons in their excited LL's are quite different
(QE and LL$_1$ in Fig.~\ref{fig-11}a,b).
In Fig.~\ref{fig-11}(c) we plot the QE--QE, QE--QE$_{\rm R}$, and
QE$_{\rm R}$--QE$_{\rm R}$ pseudopotentials obtained by 
combining the short-range data from exact diagonalization and the 
electron--electron parameters at long range.
These are the effective interactions that we used in numerics at 
$\nu^*=2$.
We also removed the (artificial) discrepancy between the degeneracy 
of $0$$\uparrow$ and $0$$\downarrow$ CF LL's on a sphere by 
considering a pair of LL shells with the same angular momentum $l$.

The computation consisted of the exact diagonalization of the $V=
[V_{\uparrow\uparrow},V_{\downarrow\downarrow},V_{\uparrow\downarrow}]$
interaction hamiltonian, separately for each combination of
$N_\uparrow$ and $N_\downarrow$ (CF numbers in the two LL's)
giving $N\equiv N_\uparrow+N_\downarrow=2l+1$, i.e., corresponding 
to $\nu^*=2$.
Note that because the length of pseudospin is not conserved by $V$, 
a separate diagonalization is required for each pseudospin projection 
$S_z=(N_\uparrow-N_\downarrow)/2$, corresponding to different 
polarizations $P=2S_z/N$ of the CF system [or $(1+P)/2$ of the whole 
electron state].

\begin{figure}
\resizebox{3.4in}{1.75in}{\includegraphics{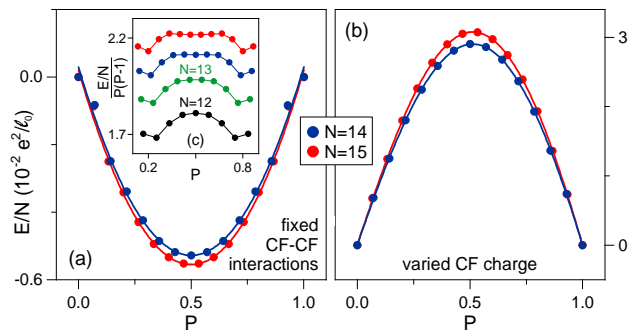}}
\caption{   (a) Energy per particle, $\varepsilon=E/N$, as a function of 
   CF spin polarization $P$, calculated for $N=14$ and 15 CF's at 
   $\nu^*=2$ using effective CF--CF interactions of Fig.~\ref{fig-11};
   curves -- parabolic fits.
   (b) Anharmonic coefficient $\alpha/4=\varepsilon/P(P-1)$ for
   $12\le N\le14$ (same units).
   (c) Same as (a) but using fill-dependent CF--CF interactions
   corresponding to a variable CF charge (see text); curves --
   polynomial fit.}
\label{fig-33}
\end{figure}

The result is the dependence of energy on polarization, $E(P)$,
tilted (by adding the appropriate linear Zeeman term $E_Z'S_z
\propto P$) to the $E(0)=E(1)$ situation.
In Fig.~\ref{fig-33}a we compare the data for two largest CF numbers
we have used, $N=14$ and 15.
For only one spin flip away from $P=0$ or 1, the ground state at
each $N$ is a single spin-wave.
For more than one spin flip, regular dependence of energy on $K/N
\sim P$ rather than on $K$ is evident.
In this ($P\sim1/2$) regime, $E$ at each $P$ scales roughly linearly 
with $N$, and the excitation energy per particle $\varepsilon(P)=
E(K/N)/N$ becomes a convergent characteristic of the macroscopic 
system.
The convergence with increasing the system size is evident, with 
a continuous $E(P)$ curve emerging for $N\rightarrow\infty$.
The data are polarization-symmetric, $E(P)=E(1-P)$, reflecting
the particle--hole symmetry at a half-filling of a pair of shells
(note that $E(0)=E(1)$ is equivalent to $\Sigma_{0\uparrow}=
\Sigma_{1\downarrow}$).

The $\varepsilon(P)$ curve determines dependence of the ground 
state polarization $P$ on the Zeeman gap in tilted-field experiments.
To calculate $P(E_Z)$, one must find the minimum of the 
total energy (per particle) including a linear Zeeman term, 
$\varepsilon(P)-P(E_Z-E_Z')/2$.
Clearly, only the convex points of $\varepsilon(P)$ can become
ground states at the appropriate $E_Z$.
A special case is a convex parabola, $\varepsilon(P)=\alpha/4
\cdot P(P-1)$, leading to a linear dependence, $P(E_Z)=1/2+
(E_Z-E_Z')/\alpha$, with $P$ varying between 0 and 1 over the 
$E_Z$ range of length $\alpha$.

Fig.~\ref{fig-33}a show that $\varepsilon(P)$ indeed is nearly 
parabolic, so in Fig.~\ref{fig-33}a we plotted $\alpha(P)=
\varepsilon(P)/P(P-1)$ to study the anharmonic contribution.
Only the result for $N=12$ ($\alpha$ having a local maximum
at $P=1/2$) agrees with the earlier calculation \cite{apalkov:02:2001}, 
also showing a downward cusp of $\varepsilon(P)$ at the 
half-polarization.
Such cusp would lead to an inflexion or a plateau in $P(E_Z)$ 
around $E_Z'$.

The emergence of a plateau would imply that the system is not 
affected by infinitesimal variation of the gap $\Delta$, i.e.,
that it is incompressible and should exhibit quantum Hall effect.
However, our calculations for larger systems seem to invalidate 
the prediction of a plateau, showing disappearance of the downward 
cusp in $\varepsilon(P)$ for $N>12$.
In Fig.~\ref{fig-33}c this is seen as transition from a local 
maximum to a local minimum in $\alpha(P)$ at $P=1/2$.
Remarkably, in experiment, the partially polarized states were 
only observed over a narrow polarization range around $P=1/2$, 
implying a well developed plateau in $P(E_Z)$, in disagreement
with the CF calculation.

\subsection{Composite fermions with fill-dependent charge}

We found notable qualitative disagreement between the numerical
results obtained (i) in terms of CF's at $\nu^*=2$ and (ii) in terms
of electrons at $\nu=2/3$ or $2/5$.  The first approach allows for
studying fairly large systems and is free of the troubling artificial
$g_0^*\ne g_1^*$ asymmetry on a sphere.  However, the results obtained
using the latter, more direct approach appear more consistent in both
used geometries.  The main conclusion, too, seems established despite
finite-size effects: the absence of a ground state of intermediate
polarization between $P=0$ and 1.
While the experiment \cite{kukushkin:05:1999} indicates a stable 
half-polarized quantum Hall state in apparent contradiction 
with approach (i), the (suggested earlier \cite{apalkov:02:2001}) 
agreement with approach (ii) is also not convincing in view 
of our numerics for larger systems and different geometries.

The most questionable assumption in using the CF model is that 
the interactions among the CF's at $\nu^*=2$ can be described
by a set of three two-body pseudopotentials, independent of 
the filling of $0$$\uparrow$ and $0$$\downarrow$ CF LL's.
Consequently, these pseudopotentials are estimated at $\nu=1/3$,
for the QE--QE, QE--QE$_{\rm R}$, and QE$_{\rm R}$--QE$_{\rm R}$
pairs (i.e., with only two CF's present in the $0$$\uparrow$ or 
$0$$\downarrow$ LL).
Such approach was proven successful only for polarized systems 
with QE fillings merely up to $\nu_{\rm QE}=1/3$ (corresponding 
to $1/3\le\nu\le4/11$) \cite{Wojs04}.
On the other hand it is well-known that the form of an actual 
electron excitation represented by a CF depends on the filling
factor.
For example, charge of a Laughlin QE at $\nu=1/3$ is $-e/3$,
while charge of QH at $\nu=2/5$ is only $e/5$.

This implies (significant) reduction of all three CF--CF 
pseudopotentials when going from $\nu=1/3$ to $2/5$, demonstrated
earlier for polarized systems \cite{Wojs00b,Lee01}.
Clearly, the pseudopotentials determined at $\nu=1/3$ cannot
be used at $\nu=2/5$ with great confidence (note, however, that
we have checked that the results are quite insensitive to the
model $V$'s used, as long as they retain qualitative behavior
at short range).
But more importantly, it probably also invalidates the concept 
of using fixed two-body pseudopotentials $V_{\uparrow\uparrow}$,
$V_{\downarrow\downarrow}$, and $V_{\uparrow\downarrow}$, which
are independent of the filling of each of the two CF LL's (at 
least in the whole range between the empty and full shell).
While the electron system at $\nu=2/3$ or $2/5$ may well be 
correctly represented by a two-pseudospin fluid of CF's with
two-body forces, the polarization-dependence of the effective 
CF--CF pseudopotentials must probably be taken into account
when modelling the $P\sim1/2$ regime.

As a test, we allowed for a very simple dependence of the CF--CF 
interactions on $P$.
We assummed a linear dependence of the charge $q_\sigma$ carried 
by a CF with pseudospin $\sigma=\,\uparrow$ or $\downarrow$ on the 
partial filling $\nu_\sigma$ of its LL (with $q_\sigma=1/3$ and
$1/5$ at $N=0$ and $g^*$, respectively).
For interaction pseudopotentials we took $V_{\sigma\sigma'}^P
=q_\sigma q_{\sigma'} V_{\sigma\sigma'}$, with $V_{\sigma\sigma'}$
shown in Fig.~\ref{fig-11}c.
The assumption that only the scale of $V$ depends on the LL filling 
(with the structure unaffected) is justified by the comparison 
of $V$ in polarized $\nu=1/3$ and $2/5$ states \cite{Wojs00b}.

The $E(K)$ calculated in this way and plotted in Fig.~\ref{fig-33}b
shows opposite (concave vs.\ convex) behavior to Fig.~\ref{fig-33}a 
obtained ignoring fill-dependence of $V$.
Again, the values at only a few spin flips away from $P=0$ or 1 scale 
best with $N$ and $K$, but a convergent $\varepsilon(P)=E(K/N)/N$ 
curve emerges around $P=1/2$.
Note that though the CF particle-hole pairs become charged for 
$q_\uparrow\ne q_\downarrow$, this artefact does not affect the 
interesting regimes of $P=0$ or 1 (exactly) or $P\sim1/2$.

The contrast between $E(K)$ shown in Figures~\ref{fig-33}a~and~b is an
obvious warning that the CF--CF interactions used so far to model
$\nu^*=2$ may have been not exact enough. 
To the best of our knowledge, the fact that quasiparticles
(of e.g. $\nu=2/5$) around $P=0$ and around $P=1$ might have
different charge and hence different interactions, has not
been previously considered. Although the
exact particular form of the interaction, $V_{\sigma\sigma'}^P$,
chosen in Fig.~\ref{fig-33}b, does not describe correctly $\nu^*=2$ in
the whole range of $P$, it does show that models of CF--CF
interactions used in earlier works (assuming all quasiparticles to
have the same charge for all values of $P$) may have given
qualitatively incorrect predictions. In order to determine whether
$E(\frac{1}{2})$ is larger or smaller than $E(0)=E(1)$, correct
CF--CF interaction must be found.

\section{Polarization one half}

Owing to the $Z_2$--symmetry of the ground state (GS), it is customary to
call the systems at $\nu=2$ an Ising--type quantum Hall
ferromagnet\cite{rezayi:05:2003} (QHF). 
Regarding only the ground state, the $\nu^*=2$
($\nu=\tf,\tt$) systems fall into the same category. Despite this, there are
substantial differences between the integer and fractional
systems. Most importantly, the possible onset of domain formation in
the integer systems is replaced by an antiferromagnetic ordering in
the systems at fractional filling.

\begin{figure}
\begin{center}
\begin{tabular}{cc}
\includegraphics[scale=.7]{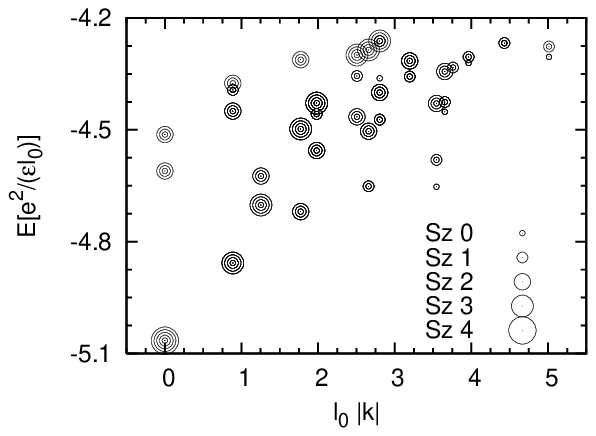}&
\includegraphics[scale=.7]{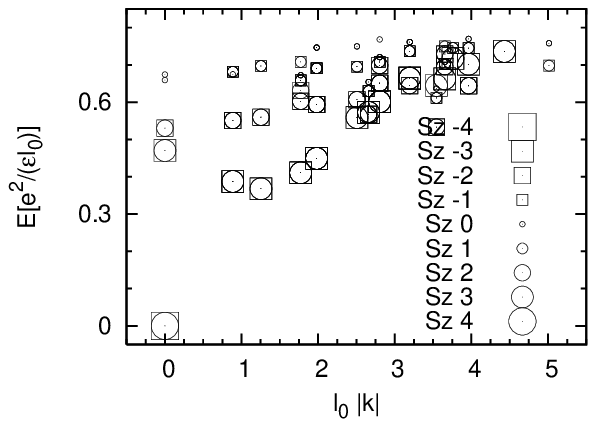}\\
(a) & (b) \\
\end{tabular}
\end{center}
\caption{Spectra of (a) $\nu=1$ and (b) $\nu=2$ quantum Hall
  ferromagnets. Eight electrons on a torus. The complete
  degeneracy of $S_z=-S,-S+1,\ldots, S$ levels (a) corresponds to a
  Heisenberg ferromagnet, the degeneracy of $S_z$ and $-S_z$ only (b) is
  proper to an Ising ferromagnet.}
\label{fig-05}
\end{figure}

When the two ferromagnetic GS's are degenerate, the complete
spectrum of the $\nu=2$ system, Fig. \ref{fig-05}b, is symmetric under
the spin inversion. While Heisenberg ferromagnets, have a spin-wave
with vanishing energy at $k\to 0$ as the lowest excitation,
Fig. \ref{fig-05}a, the first excited state of a $\nu=2$ system has a
single spin flip at a finite wavevector $k$, Fig. \ref{fig-05}b. This
is in line with Goldstone theorem which requires a continuous symmetry
of the GS, which is SU(2) in the former case. 

\begin{figure}
\begin{center}
\begin{tabular}{cc}
$2/3$ (a) & $2/5$ (b) \\
\includegraphics[angle=0,scale=0.4]{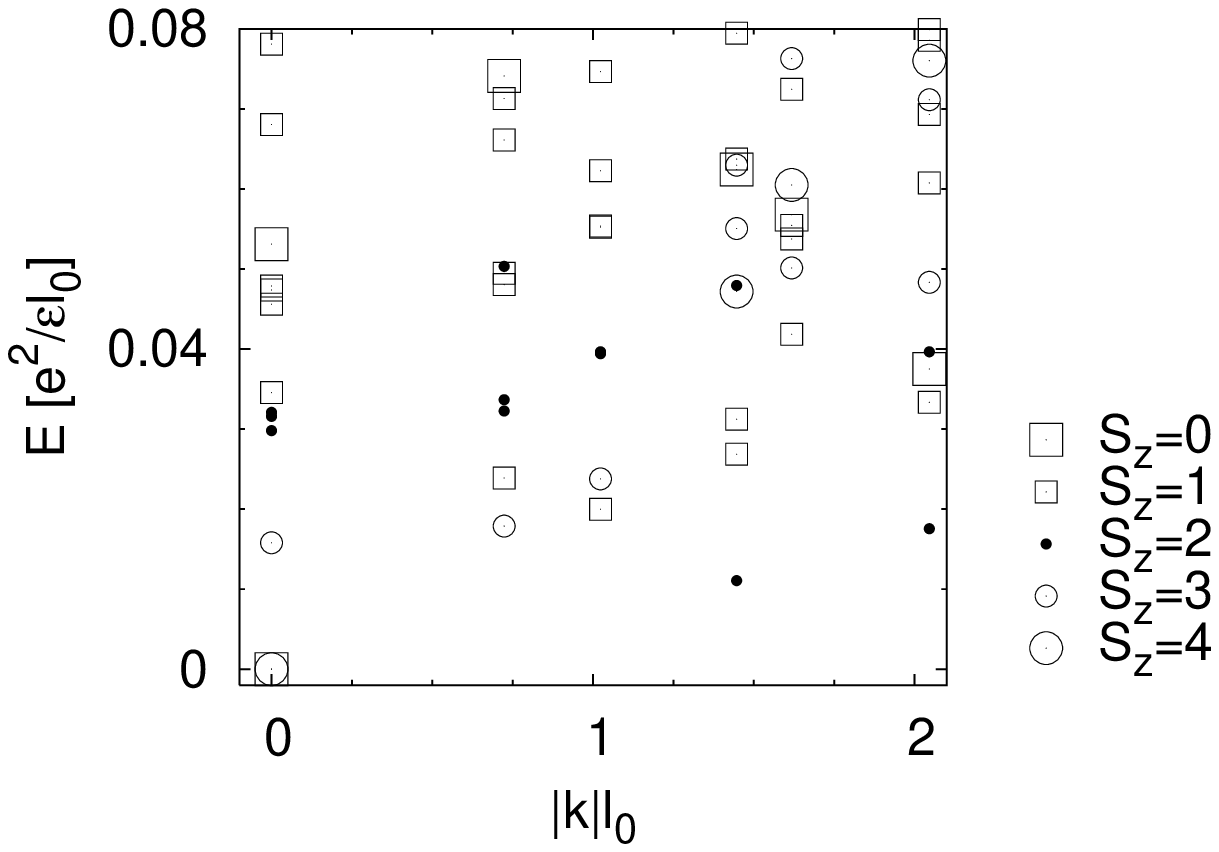} &
\hskip-.9cm\includegraphics[angle=0,scale=0.4]{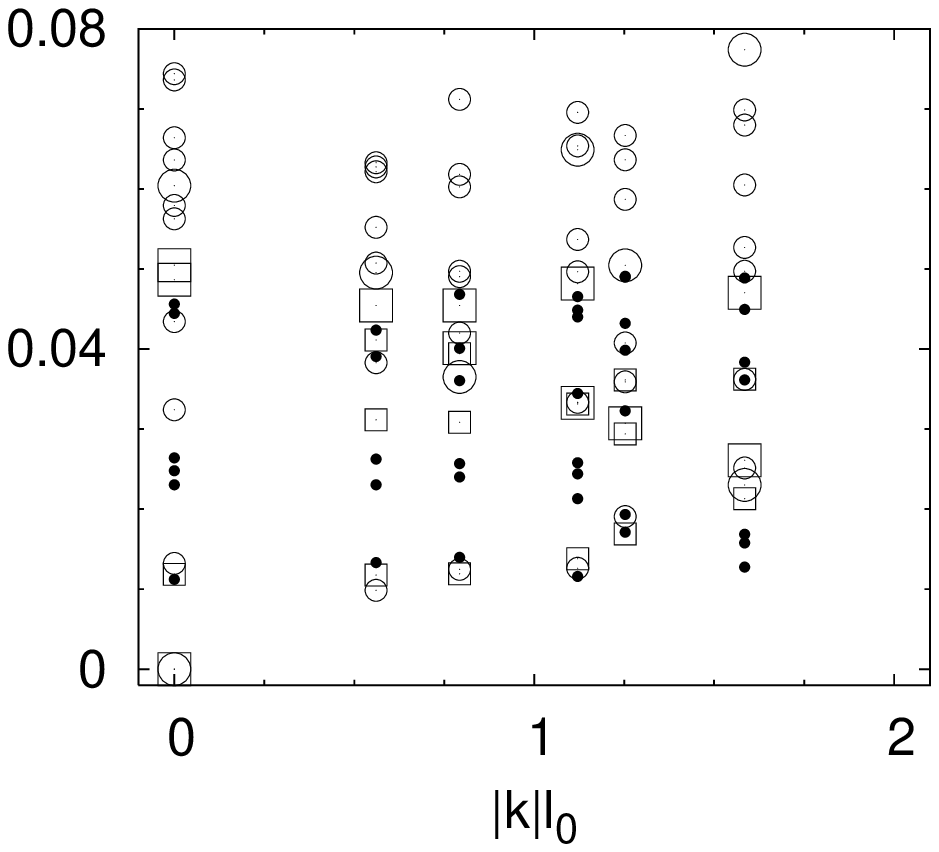}
\end{tabular}
\end{center}
\caption{Full spectrum of eight electrons at $\nu=\tt$ and $\tf$
  (hence both $\nu^*=2$) on a torus with Zeeman energy adjusted so
  that the fully polarized and spin-singlet incompressible states are
  degenerate. To be compared with $\nu=2$, Fig. \ref{fig-05}b. }
\label{fig-06}
\end{figure}

The corresponding full spectra of the $\nu^*=2$ systems,
Fig. \ref{fig-06}, do not have any obvious structure resembling the
one of $\nu=2$, Fig. \ref{fig-05}b. The spin--inversion symmetry is
missing, $S_z=3.0$ and $S_z=1.0$ states have different energies,
Fig. \ref{fig-06}. No definite prediction can be made
about the spin of the lowest excitation.
Perhaps most importantly, the $\nu=\tt$ and $\tf$ spectra look very
differently, Fig. \ref{fig-06}a,b, 
except for the GS at $k=0$ and its gap. This is markedly
at odds with the picture of non-interacting composite
fermions according to which the spectra should be the same after
rescaling to equal effective magnetic length.

\begin{figure}
\begin{center}
\unitlength=1mm
\begin{tabular}{cc}
\includegraphics[scale=.6]{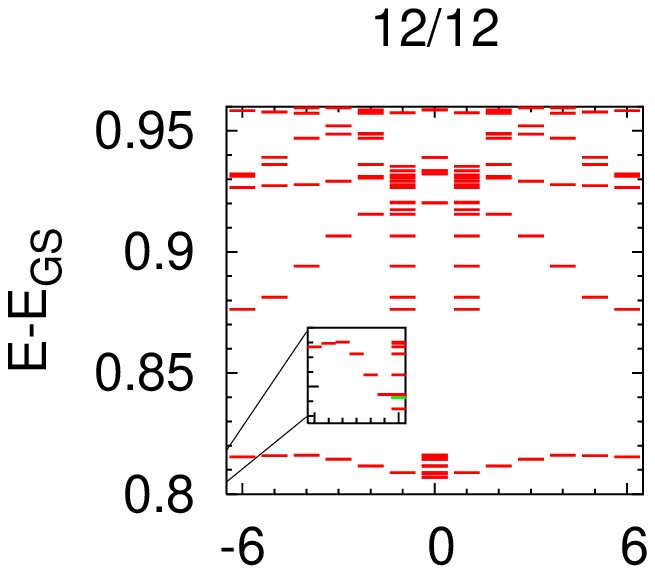} & 
\hskip-1cm \includegraphics[scale=.6]{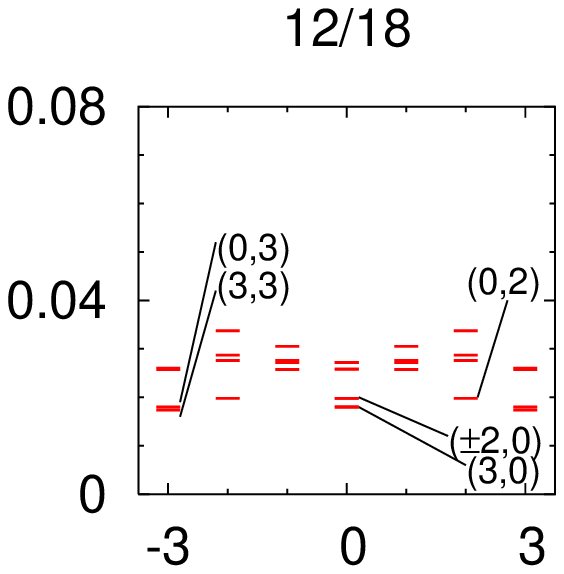} \\[-3mm]
\hskip1cm (a) & (c) \\[3mm]
\includegraphics[scale=.6]{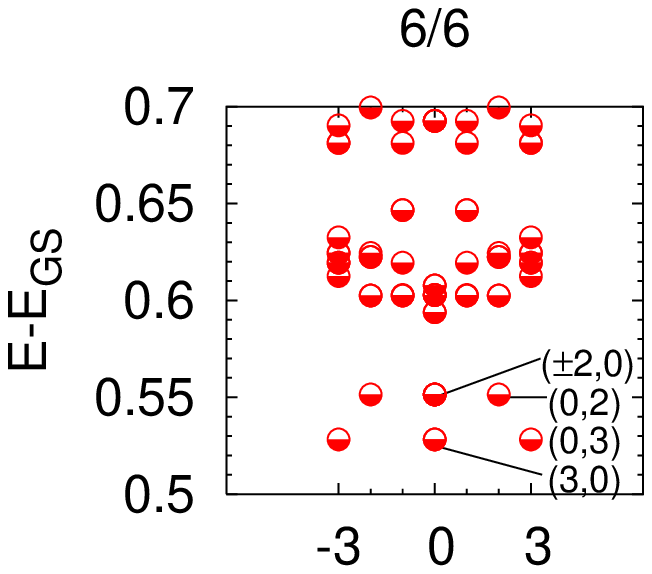}%
\put(-26,2){$k_y\cdot N_m a/\pi$}
 & \hskip-1cm \includegraphics[scale=.6]{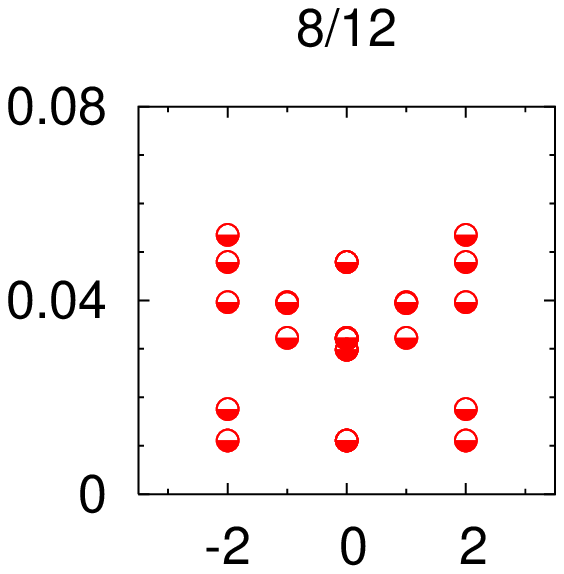}%
\put(-26,2){$k_y\cdot N_m a/\pi$} \\
\hskip1cm (b) & (d) \\
\end{tabular}
\end{center}
\caption{Spectra of the half--polarized sector  in a $\tt$
  and $\nu=2$ systems. The titles give $N/N_m$, number of electrons
  $N$ and filling factor $\nu=N/N_m$. The reference energy is the
  ferromagnetic ground state at degeneracy. Small numbers at some of
  the states are their linear momenta $(k_x,k_y)/k_u$.
}
\label{fig-07}
\end{figure}

Let us now concentrate on the sector of $S_z$ corresponding to equal number
of up and down spins in the active Landau levels. For $\nu=\tt,\tf$ using the 
ED with electrons this means $S_z=N/4$ and for $\nu=2$ with the
low--lying $0$$\dn$ level neglected, it is $S_z=0$. 


The spectrum of a $\nu=2$ system has a clear structure, Fig. \ref{fig-07}a.
A $(2N-2)$--tuplet of states distinguished by
\begin{equation}\label{eq-01}
\krv/k_u=(0,\pm i)\mbox{ or }(\pm i,0)\,,\quad i=0,1,2,\ldots, N/2,
\end{equation}
is separated from higher excited states. Rezayi
\etal\cite{rezayi:05:2003} identified this group as a state with two
(Ising) domains in different system of the same type.

\begin{figure}
\begin{center}
\begin{tabular}{cc}
\includegraphics[angle=0,scale=.4]{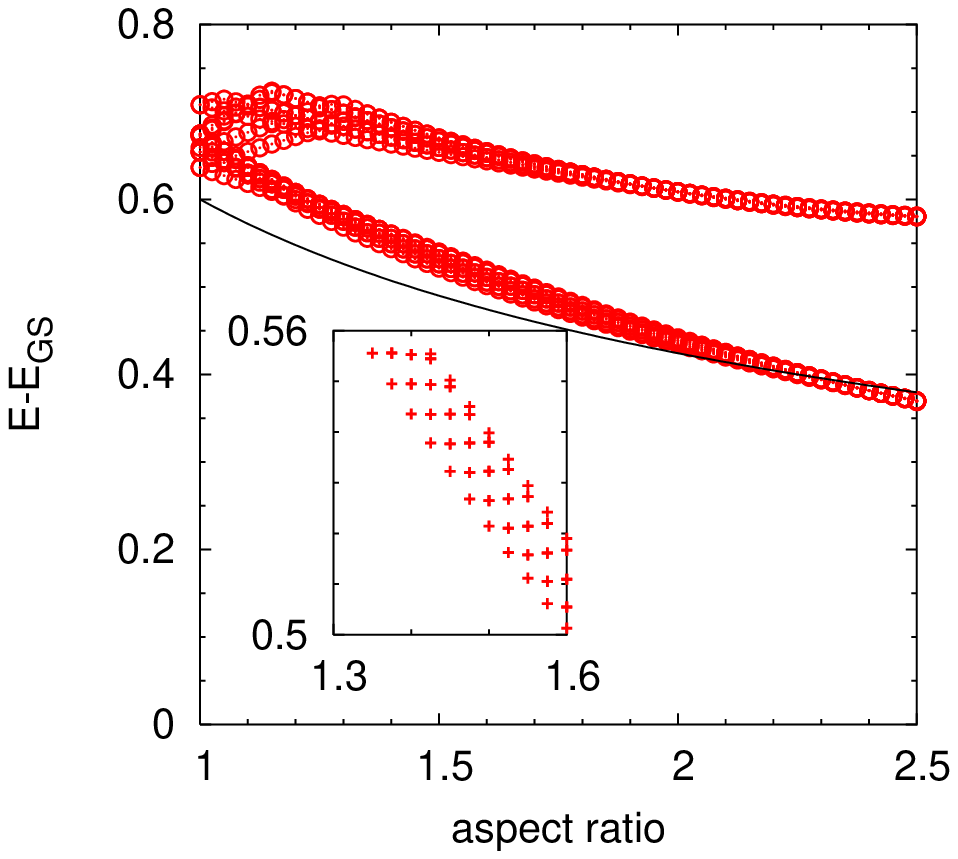} &
\hskip-1cm\includegraphics[angle=0,scale=.4]{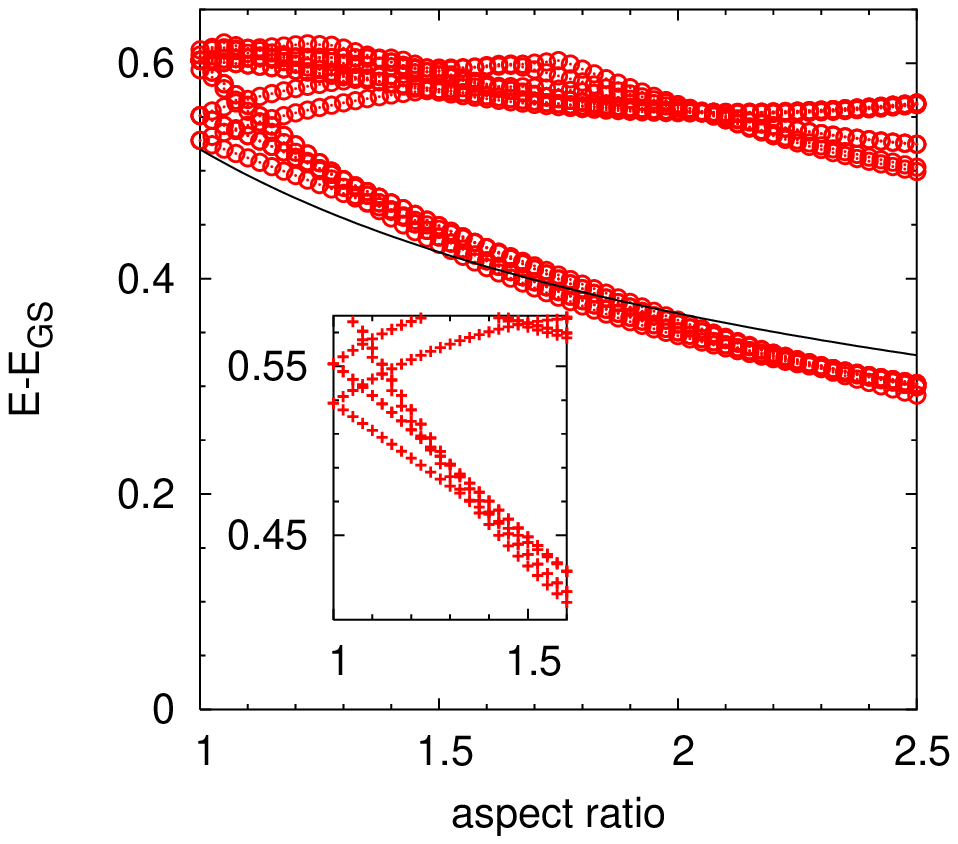} \\
(a) & (b)
\end{tabular}
\end{center}
\caption{Energies of the $S_z=0$ states in the $\nu=2$ Ising
  ferromagnet relative to the ground state ($S_z=\pm N/2$). The solid
  line is a fit $c/\sqrt{\alpha}$, $\sqrt{\alpha}\propto b$. Right
  $N=6$ ($c=0.52$), left $N=8$ ($c=0.60$). }
\label{fig-10}
\end{figure}

The easiest way to see this is in the spin-resolved density-density
correlation functions\cite{vyborny:09:2006}. Another
possibility\cite{rezayi:05:2003} is to replace the square in our model by a
rectangle with periodic boundary conditions while keeping its area
fixed, Fig.~\ref{fig-10}. A group of $N$ states quickly detaches from the
$(2N-2)$-tuplet, once the aspect ratio $\alpha=a:b$ of the rectangle
exceeds $\approx 1.2$. These states have $\krv/k_u=(\pm i,0)$, 
$i=0,1,2,\ldots, N/2$, their degeneracy improves with
increasing $\alpha$ and the energy changes roughly proportional to
$1/\sqrt{\alpha}\propto b$. It is very suggestive that the change of
energy is mostly due to changing length of the domain walls which are
likely to be oriented along the shorter side of the rectangle.

This method of investigation is particularly useful for small
systems. It is not possible to distinguish the
$(2N-2)$-tuplet in a $N=6$ system ($\nu=2$), because its degeneracy is
far from being perfect, so that it is mixed up with higher excited states,
Fig.~\ref{fig-07}b. A minute variation of the aspect ratio, however, 
separates the $N$-tuplet of states with domains oriented parallel to $b$, 
Fig.~\ref{fig-10}a. The energy cost of a domain wall per magnetic
length obtained for both system sizes in Fig.~\ref{fig-10} is the
same, $0.042\cunit$.

\begin{figure}
\begin{center}
\begin{tabular}{cc}
\includegraphics[scale=.5]{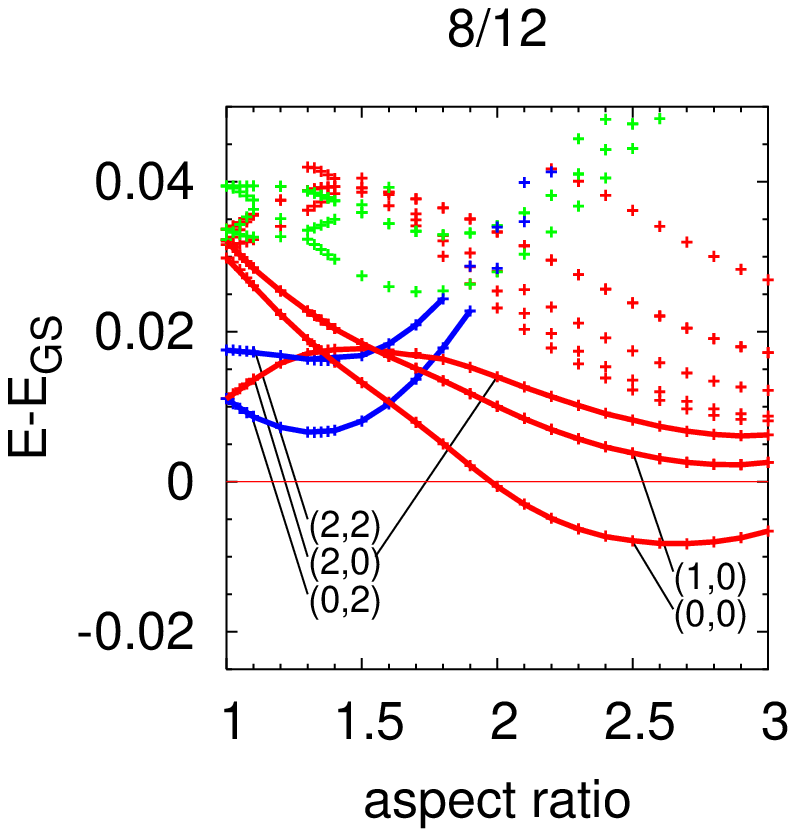} &
\includegraphics[scale=.5]{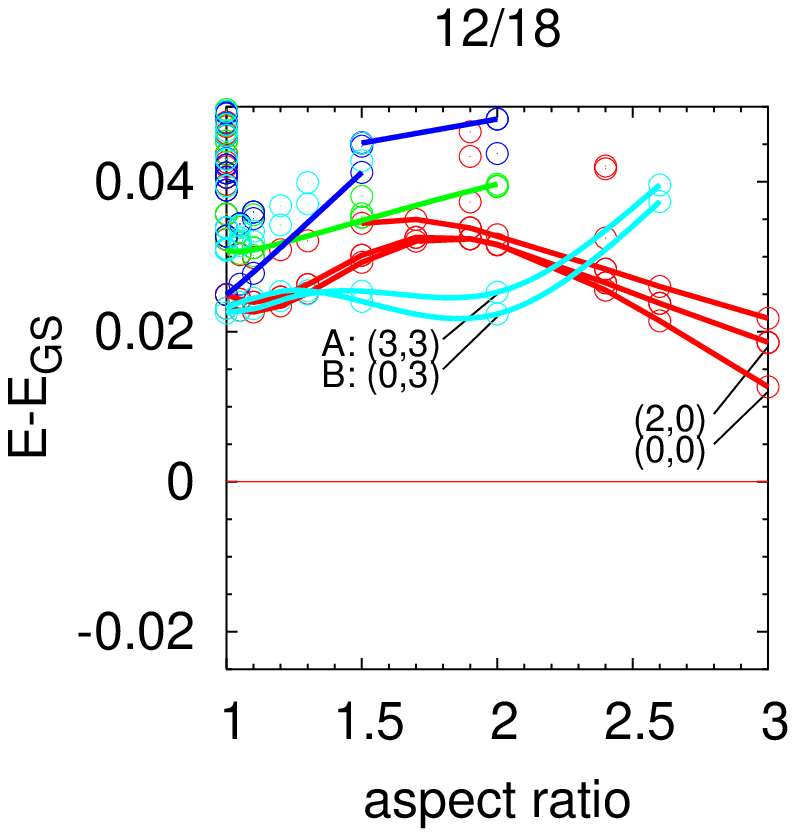} \\
(a) & (b)\\
\end{tabular}
\end{center}
\caption{Energies of the HPS under varying aspect ratio (Coulomb
  interaction, $\tt$). (a) $N=8$, (b) $N=12$. 
}
\label{fig-08}
\end{figure}

The $\nu=\tt$ system has again a rather different spectrum of the
$S_z=N/4$ sector, Fig.~\ref{fig-07}c,d compared to
Fig.~\ref{fig-07}a,b. No similar grouping of states is obvious. On the
other hand, because six electrons should be in the inactive CF LL in
the $N=12$ system for instance, we should also keep in mind a
comparison between the $(N,N_m)=(12,18)$ and $(6,6)$ spectra,
Fig.~\ref{fig-07}c and \ref{fig-07}b. In both systems, the lowest
energy states form a loose group (the marked states in
Fig.~\ref{fig-07}bc).  This group is separated from
other states by $0.01$ (Fig.~\ref{fig-07}c) and $0.04\cunit$
(Fig.~\ref{fig-07}b) what is slightly more than differences
between energies within the group. The states can be classified by
their momentum $\krv/k_u$: these are $(\pm 2,0),(3,0)$ for $\nu=2$ and
$(\pm 2,0),(3,0),(3,3)$ for $\nu=\tt$ plus their $x-y$ symmetric
states. For $\nu=2$, these states belong to the $(2N-2)$--tuplet of the
single--domain state.

The $\krv/k_u=(3,3)$ state of $\nu=\tt$, Fig.~\ref{fig-07}c, as the
only clear difference between Fig.~\ref{fig-07}c and b, cannot be just 
a finite size artefact.  Under
a slight squeeze, the $\tt$ systems reveal a clearly
different behaviour compared to $\nu=2$. The $\krv/k_u=(3,3)$ state
(marked by A in Fig.~\ref{fig-08}b) quickly becomes the absolute
ground state of the system, together with the $\krv/k_u=(0,3)$ state
(B in Fig.~\ref{fig-08}b). These two states react to the squeezing very
similarly within the range $1.3<\alpha<2.3$.
With some experience from $\nu=2$
systems, Fig.~\ref{fig-10}, this range of $\alpha$ may correspond to
the lifting of the $x-y$ degeneracy while still preserving the 2D character
of the system ($\alpha$ not too far from one).  An $N$-tuplet similar
to the integer filling systems (Eq. \ref{eq-01}) does not appear as
far as for
$\alpha<3$. Even though such grouping is possible for larger aspect
ratios, their eventual relevance would have to be supported by some
strong external anisotropy justifying the large aspect ratio chosen
for the model.

Going from smaller to larger systems, Fig.~\ref{fig-08}a
and~\ref{fig-08}b, it seems indeed possible that the two states,
$\krv/k_u=(0,N/2), (N/2,N/2)$ become the lowest states with $S_z=N/4$
when the $x-y$ symmetry is lifted. Namely,
$\krv/k_u=(0,2),(2,2)$ for $N=8$ and $\krv/k_u=(0,3),(3,3)$ for $N=12$
do and $k_x/k_u=2$ and $3$ are the maximal $k_x$-values
in the finite system with $N=8$ and $12$ electrons.

\begin{figure}
\begin{center}
\begin{tabular}{ccc}
(a) & (b) & (c) \\
\hbox{\includegraphics[angle=0,scale=.35]{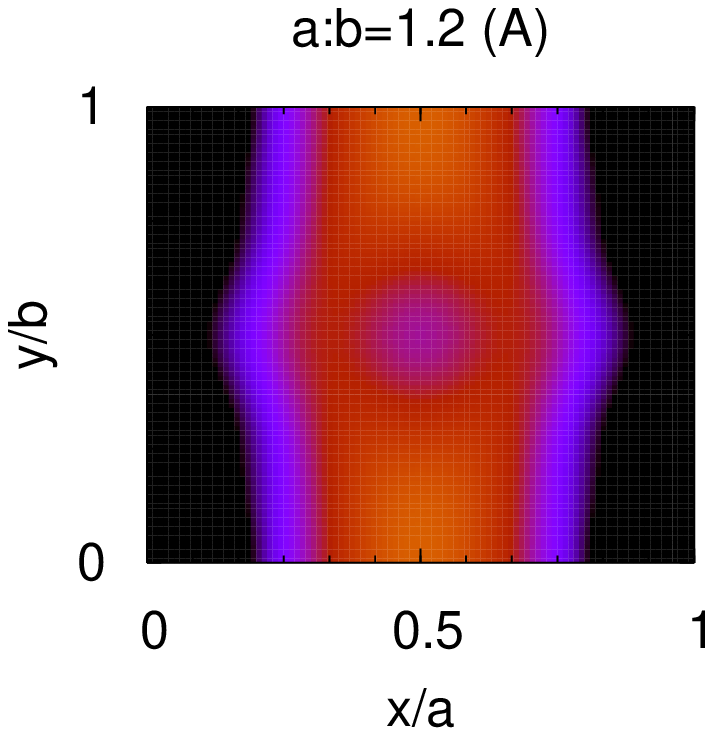}} &
\raise4mm\hbox{\includegraphics[angle=0,scale=.35]{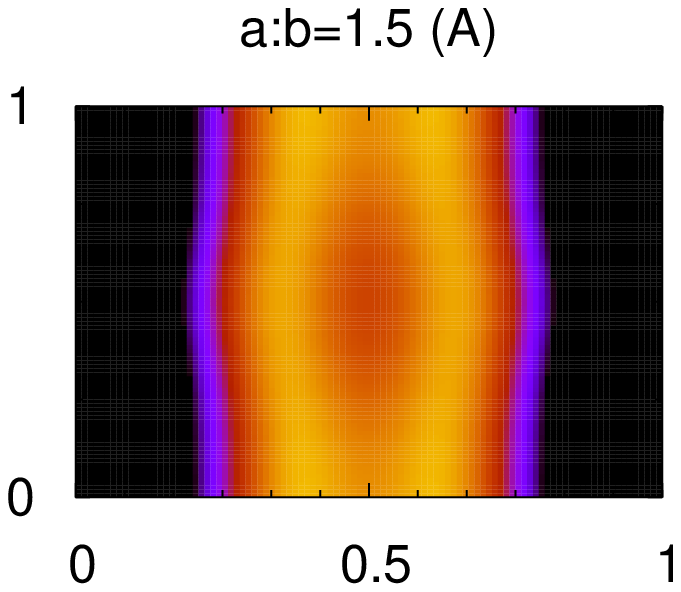}}&
\raise8mm\hbox{\includegraphics[angle=0,scale=.35]{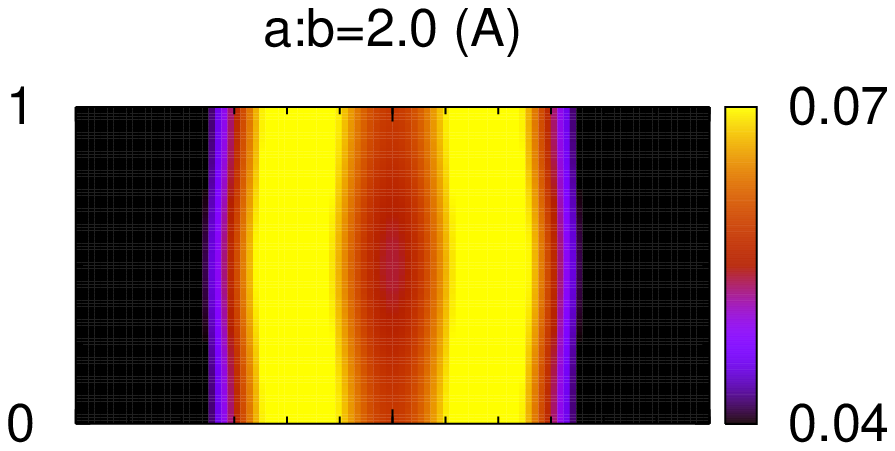}}\\ 
\span{\raise3cm\hbox{(d)} \includegraphics[angle=0,scale=.35]{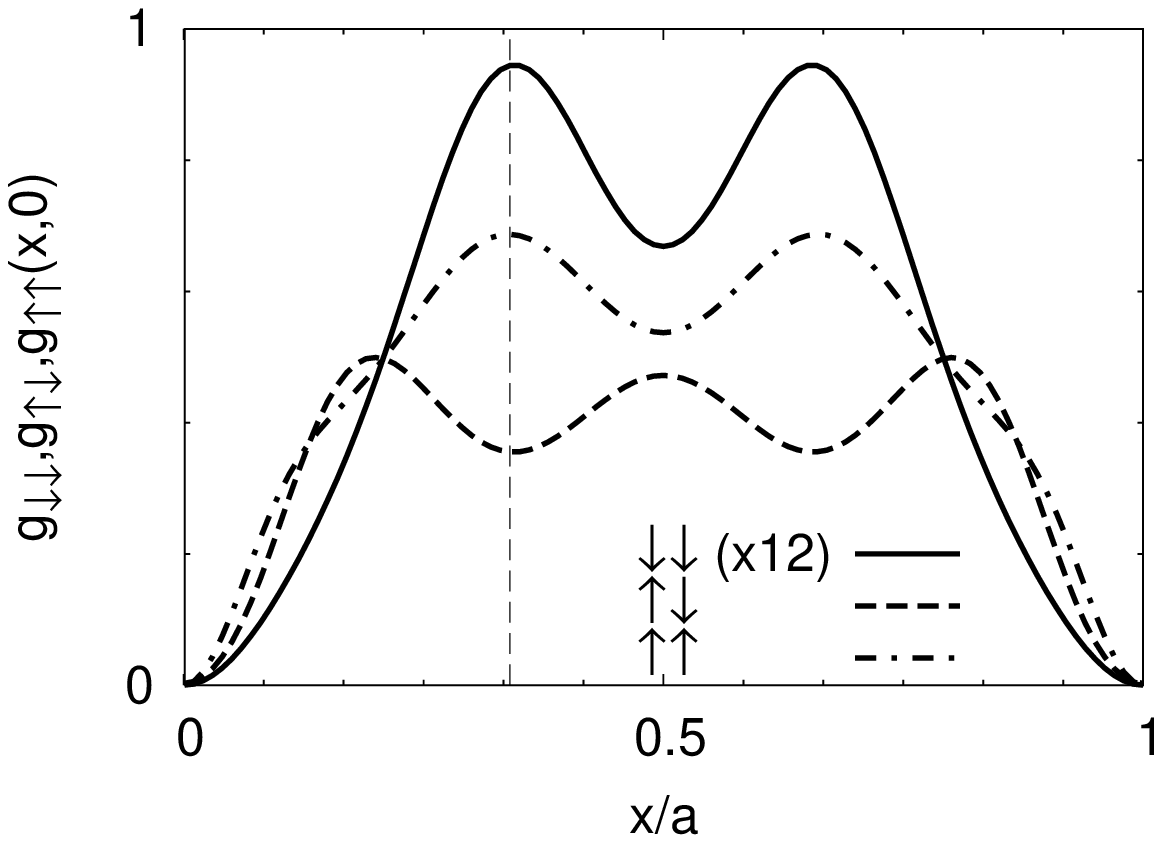}} &
\raise18mm\hbox{\includegraphics[angle=0,scale=.35]{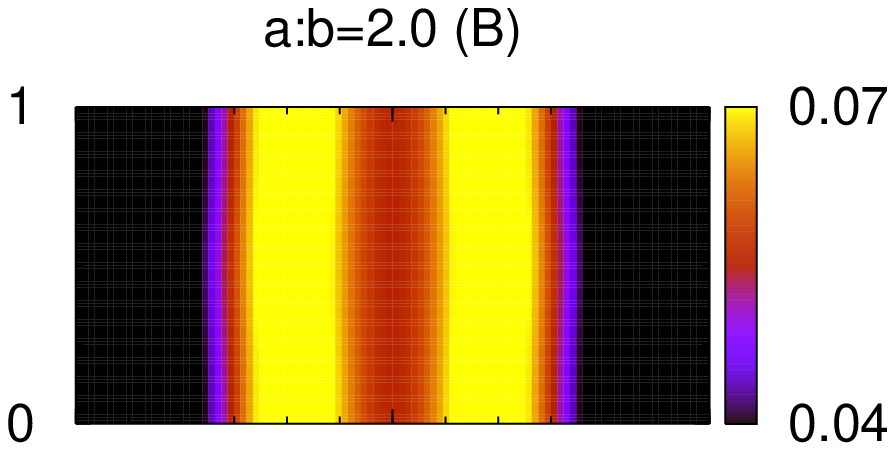}}
\unitlength=1mm
\put(-20,10){(e)}
\end{tabular}
\end{center}
\caption{Correlation functions $g(\vek{r})=g(x,y)$ of the
  $\krv/k_u=(3,3)$ half-polarized state (A in
  Fig.~\ref{fig-08}b). Under slight variation of the aspect ratio
  $\alpha=a:b$, the conditional probability $g_{\up\up}(\vek{r})$
  indicates antiferromagnetic ordering (a,b,c), confirmed by
  $g_{\dn\dn}$ and $g_{\up\dn}$, (d). The state $B$ from
  Fig.~\ref{fig-08}b has a very similar structure (e).}
\label{fig-09}
\end{figure}

Albeit distinguished by $\krv$, the two states (A,B in
Fig.~\ref{fig-08}b) look very similar in their spin--resolved
density--density correlation functions 
$$g_{\up\up}(\vek{r})=\langle \delta(\vek{r}_1-\vek{r}_2-\vek{r})
\delta_{\sigma_1\up}\delta_{\sigma_2\up}\rangle\,,
$$
as shown in Fig.~\ref{fig-09}c,e. Other combinations of spins not shown in 
Fig.~\ref{fig-09} ($\dn\dn$, $\up\dn$) also
confirm this conclusion. We observe for both states two vertical
stripes (maxima) in $g_{\up\up}(\vek{r})$, $g_{\dn\dn}(\vek{r})$
together with the two complementary stripes (minima) in
$g_{\up\dn}(\vek{r})$, Fig.~\ref{fig-09}d. This could mean that six
electrons in the active CF Landau levels align antiferromagnetically,
$\up\dn\up\dn\up\dn$ following the elongated side of the elementary
cell.

\section{Discussion and summary}

With near degeneracy of $0$$\downarrow$ and $1$$\uparrow$ CF LL's,
quantum Hall ferromagnet of $\nu^*=2$ is a unique system: (i) two
different incompressible CF liquids form for P=0 and 1; (ii)
low-energy excitations of both liquids involve spin; (iii) experiment
suggests another liquid at P=1/2; (iv) in the P=1/2 liquid
interactions among CF's play crucial role (unlike at P=0 or 1 where
incompressibility is due to LL filling and the interactions are not
important). All together, (v) the concept of incompressible states of
correlated CF's at these fillings appears even more fascinating than
in the states discovered more recently at $\nu=4/11$ or $3/8$
\cite{Pan03}, due to additional spin freedom.


The comparison of the experimental evidence with our various numerical 
calculations demonstrates the lack of understanding of the microscopic 
origin of the half-polarized quantum Hall states.
Yet, it appears very difficult to model these states in finite-size
numerics.

In the electron calculation, large Hamiltonian dimensions make exact 
calculation of the $P=0$ state very complicated already for $N>8$.
For $N=8$ only two spin flips separate $P=0$ or 1 from $P=1/2$ 
(the first one being simply a spin-wave), which might not be enough 
to capture physics of the correlated $P\sim1/2$ regime.
Moreover, calculations on a sphere suffer from the $g_0^*\ne g_1^*$ 
artefact that further complicates interpretation of the results
in this geometry.

In the CF calculation, the result strongly depends on the choice 
of effective CF--CF interactions, which are not known with near 
enough accuracy.
This problem does not appear in the understanding of Jain states
corresponding to filled CF LL's (provided these interactions are 
weaker than $\hbar\omega_c^*$), but here it is essential.
In the CF picture, one particle or hole in a CF LL represents 
different electronic excitations depending on the filling of 
a CF shell. 
These excitations are only known in some special cases, when 
they correspond to e.g., Laughlin QE's or QE$_{\rm R}$'s (in 
an empty CF LL) or Jain QH's (in a full CF LL).
Consequently, although it seems plausible that the low-energy 
dynamics of the electron states corresponding to partially filled 
CF LL's is generally well described by two-body effective CF--CF 
interactions, their pseudopotentials are not well known.

It seems that electron calculations in larger systems (preferably in
toroidal geometry) are needed for understanding of the occurrence and
incompressibility of half-polarized $\nu=2/5$ and $2/3$ states.  More
advanced exact diagonalization as well as Monte Carlo methods must be
considered.  On the other hand, further experimental studies are much
needed in view of possible insight into the nature of CF--CF
interactions.

In summary, at the level of present computational capacity, the
calculations for toroidal {\em and} spherical geometry in Sec. IIA
indicate that the ferromagnet-paramagnet transitions both at $\nu=2/3$
and $2/5$ are abrupt. This applies to homogeneous and isotropic
systems. The antiferromagnetically ordered states at polarization one
half (Sec. III) could in principle, however, become the absolute
ground state near the transition if a suitable anisotropy or
inhomogeneity in the system is present. These states also constitute
probably the best demonstration of marked differences between the
fractional and integer QHF. In the latter case, the system splits
into two equally large domains ($\up\up\up\dn\dn\dn$).

The authors acknowledge support from the following grants:
AV0Z10100521 of the Academy of Sciences of the Czech Republic
(KV and O\v C),
LC510 of the Ministry of Education of the Czech Republic --
Center for Fundamental Research (KV and O\v C),
2P03B02424 of the Polish MNiSW (DW and AW),
and DE-FG 02-97ER45657 of US DOE (JJQ).


\begin{thebibliography}{99}
\bibitem{comm:02}
$[N/2]$ denotes the whole part of $N/2$. For $N$ even, the momenta
  $k_x/k_u=\pm N/2$ are indistinguishable due to the PBC.
\end{thebibliography}

\end{document}